\documentclass[12pt]{iopart}

\usepackage{graphicx}
\usepackage{adjustbox}
\usepackage{aas_macros}

\usepackage{hyperref}
\usepackage{bm}
\usepackage{color}

\usepackage{caption}
\usepackage{subcaption}

\begin{document}

\title[Bayesian search for gravitational wave bursts in PTA data]{Bayesian search for gravitational wave bursts in pulsar timing array data}

\author{Bence B\'ecsy, Neil J.~Cornish}

\address{eXtreme Gravity Institute, Department of Physics, Montana State University, Bozeman, Montana 59717, USA}
\ead{bencebecsy@montana.edu}
\vspace{10pt}
\begin{indented}
\item[]April 2021
\end{indented}

\begin{abstract}
The nanohertz frequency band explored by pulsar timing arrays provides a unique discovery space for gravitational wave signals. In addition to signals from anticipated sources, such as those from supermassive black hole binaries, some previously unimagined sources may emit transient gravitational waves (a.k.a. bursts) with unknown morphology. Unmodeled transients are not currently searched for in this frequency band, and they require different techniques from those currently employed. Possible sources of such gravitational wave bursts in the nanohertz regime are parabolic encounters of supermassive black holes, cosmic string cusps and kinks, or other, as-yet-unknown phenomena. In this paper we present \texttt{BayesHopperBurst}, a Bayesian search algorithm capable of identifying generic gravitational wave bursts by modeling both coherent and incoherent transients as a sum of Morlet-Gabor wavelets. A trans-dimensional Reversible Jump Markov Chain Monte Carlo sampler is used to select the number of wavelets best describing the data. We test \texttt{BayesHopperBurst} on various simulated datasets including different combinations of signals and noise transients. Its capability to run on real data is demonstrated by analyzing data of the pulsar B1855+09 from the NANOGrav 9-year dataset. Based on a simulated dataset resembling the NANOGrav 12.5-year data release, we predict that at our most sensitive time-frequency location we will be able to probe gravitational wave bursts with a root-sum-squared amplitude higher than $\sim 5 \times 10^{-11}$ Hz$^{-1/2}$, which corresponds to $\sim 40 M_{\odot} c^2$ emitted in GWs at a fiducial distance of 100 Mpc.
\end{abstract}

%
%
%
%
%

\section{Introduction}
\label{sec:intro}

The detection of nanohertz gravitational waves (GWs) is the main objective of pulsar timing arrays (PTAs). These large scale experiments regularly monitor a collection of millisecond pulsars to achieve this goal (see e.g.~\cite{pta_review}). The three major pulsar timing arrays currently operating are the North American Nanohertz Observatory for Gravitational Waves (NANOGrav,~\cite{nanograv_12p5yr_data}), the European Pulsar Timing Array (EPTA,~\cite{epta_data}), and the Parkes Pulsar Timing Array (PPTA,~\cite{ppta}). In addition, there are emerging PTA efforts in India (InPTA,~\cite{InPTA}), China (CPTA,~\cite{CPTA}), and South Africa (SAPTA,~\cite{SAPTA}). These project are collaborating under the International Pulsar Timing Array (IPTA,~\cite{ipta, ipta_dr1, ipta_dr2}) consortium.

The most promising GW sources in the nHz regime are supermassive black hole binaries (SMBHBs). These can be detected through observing their collective effect as a red noise process with quadrupolar correlations between pulsars (see e.g.~\cite{nanograv_12p5yr_gwb}). The most massive and nearby of these SMBHBs might also be detectable individually (see e.g.~\cite{nanograv_11yr_CW}). It has also been proposed that these two searches can be carried out simultaneously, thus accounting for any interaction between the two types of signatures~\cite{BayesHopper}. Searches have also been carried out for bursts with memory~\cite{nanograv_11yr_burst_with_memory}, i.e., a permanent deformation of spacetime after a violent astrophysical event, like the merger of two black holes.

In this paper, we focus on searching for generic GW transients (a.k.a.~GW bursts) in PTA data. The subject has been studied extensively in the context of ground-based interferometric GW detectors, where many different algorithms are in use~\cite{BayesWave, BayesLine, BayesWaveIII, cWB, STAMP-AS, oLIB} and several searches have been carried out throughout different observing runs of the LIGO~\cite{LIGO} and Virgo~\cite{Virgo} detectors~\cite{O2_all_sky_short,O2_all_sky_long}. The detection problem of GW bursts has also been considered for space-based detectors~\cite{LISA_burst_search}.

In the PTA context, no analysis of real data has been carried out so far. However, several methods have been suggested, including an analytical Bayesian framework~\cite{Finn_Lommen2010}, a Bayesian nonparametric approach~\cite{deng_pta_burst}, and a frequentist search working in the time-frequency domain~\cite{Zhu_et_al_burst}. Ref.~\cite{pulsar_terms_in_burst_search} suggests a possible improvement by considering the coherence between the pulsar terms, which appears in some cases. In this paper we present \texttt{BayesHopperBurst}\footnote{\url{https://github.com/bencebecsy/BayesHopperBurst}}, a Bayesian search for GW bursts in PTA data, which is based on a trans-dimensional Reversible Jump Markov Chain Monte Carlo (RJMCMC) sampler~\cite{rjmcmc, rjmcmc_new} akin to that used in the BayesWave algorithm~\cite{BayesWave} to search for and reconstruct GW bursts in the data of ground-based GW detectors. The work described in this paper is the continuation of that presented in~\cite{Justin_Neil_transdim_noise}, where the authors used similar methods to model noise transients in PTA data. One promising source of nHz GW bursts are parabolic (or highly eccentric) encounters of two SMBHs, which can occur in a hierarchical triple system of SMBHs~\cite{ebbh_from_triples}. Cosmic string kinks and cusps can also produce GW busts in this frequency range~\cite{Damour_et_al_cosmic_strings}.

The paper is organized as follows. In Section~\ref{sec:methods} we describe the methods employed by \texttt{BayesHopperBurst} including the model used to describe transient signals and the sampling techniques that enables it to effectively explore the parameter space. In Section~\ref{sec:injection_tests} we perform various injection tests, demonstrating that \texttt{BayesHopperBurst} recovers a wide variety of signals and noise transients. In Section~\ref{sec:upper_limit} we analyze a simulated dataset similar to the latest NANOGrav data release~\cite{nanograv_12p5yr_data} to make a prediction of what we can expect from analyzing the real dataset. In Section~\ref{sec:real_data_test} we analyze B1855+09 from the NANOGrav 9-year dataset~\cite{nanograv_9yr_data} to demonstrate our algorithm's noise transient modeling capabilities on real data. Finally, we offer concluding remarks in Section~\ref{sec:conclusion}.

\section{Methods}
\label{sec:methods}

In this section, we introduce the model used to describe signals and noise transients with arbitrary morphology, and we provide some details of the sampling techniques employed to efficiently explore the resulting high-dimensional parameter space.
We use the \texttt{enterprise}\footnote{\url{https://github.com/nanograv/enterprise}}~\cite{enterprise} software package for handling PTA data and calculating the likelihood used in our Bayesian analysis.
Timing models were produced by \texttt{libstempo}\footnote{\url{https://github.com/vallis/libstempo}}, which is a \texttt{python} wrapper for the \texttt{tempo2}\footnote{\url{https://bitbucket.org/psrsoft/tempo2/}} timing package.
We also make use of the \texttt{la\_forge}\footnote{\url{https://github.com/Hazboun6/la_forge}}~\cite{la_forge} package for some of our figures.

\subsection{Model}
\label{ssec:model}

The model we use to describe our dataset has many similarities with the one used in the \texttt{BayesHopper} algorithm. The only fundamental difference is that the collection of sinusoids are replaced by a set of ``signal'' wavelets coherent across detectors and a set of ``noise transient'' wavelets which only appear in a given pulsar and describe transient noise features in the data. In this section we give a brief description of the model and focus on differences between \texttt{BayesHopper} and \texttt{BayesHopperBurst}. More details can be found in~\cite{BayesHopper}.

Consider a PTA consisting of $N$ pulsars. The $i$th residual in the $k$th pulsar of the array is modeled as:
\begin{equation}
 \delta t_{ki} = \sum_l M_{kil} \delta \xi_{kl} + n_{ki} + g_{ki} + w_{ki}(\bm{\theta}_{\rm s}) + v_{ki}(\bm{\theta}_{\rm n}),
 \label{eq:model}
\end{equation}
where $\delta \xi_{kl}$ is the offset from the best fit value of the $l$th timing model parameter, and $M_{kil}$ is the design matrix, which represents a timing model linearized around the best fit parameter values. $n_{ki}$ represents all noise processes unique to each pulsar, while $g_{ki}$ is the contribution of an isotropic stochastic GW background (GWB) which is correlated between different pulsars. The contribution of the transient GW signal is represented by $w_{ki}(\bm{\theta}_{\rm s})$, while that of incoherent transient noise features is described by $v_{ki}(\bm{\theta}_{\rm n})$, where $\bm{\theta}_{\rm s}$ and $\bm{\theta}_{\rm n}$ represent the parameters describing the GW signal and the transient noise respectively.

\subsubsection{Transient noise model}
\label{ssec:glitch_model}

The contribution of noise transients to the timing residual ($v_{ki}(\bm{\theta}_{\rm n})$) is modeled as a sum of sine-Gaussian (Morlet-Gabor) wavelets:
\begin{equation}
 v_{ki}(\bm{\theta}_{\rm n}) = \sum_{j=1}^{M_k} \Psi(t_{ki}; \bm{\lambda}_{kj}),
\end{equation}
where $M_k$ is the number of wavelets used in the $k$th pulsar, $t_{ki}$ is the $i$th observing time of the $k$th pulsar, and $\bm{\lambda}_{kj}$ is the parameter vector describing the $j$th wavelet in the $k$th pulsar, which is related to the full parameter vector as $\bm{\theta}_{\rm n} = (M_1, \bm{\lambda}_{1 1}, ..., \bm{\lambda}_{1 M_1}; ... ; M_N, \bm{\lambda}_{N 1}, ..., \bm{\lambda}_{N M_N})$, and the wavelets are defined as:
\begin{equation}
\Psi(t_i; \bm{\lambda}_{kj}) = A e^{(t-t_0)^2/\tau^2} \cos(2\pi f_0 (t-t_0) + \phi_0),
\end{equation}
where $A$ is the amplitude, $t_0$ is the central time, $\tau$ is the characteristic duration, $f_0$ is the central time, and $\phi_0$ is the initial phase. We can see that a single wavelet can be described by 5 parameters, i.e.~$\bm{\lambda}_{kj} = (A, t_0, \tau, f_0, \phi_0)$. We use Morlet-Gabor wavelets as they have the compelling property of having the smallest time-frequency area allowed by the Heisenberg–Gabor limit~\cite{gabor1946theory}. Other functions, such as shapelets~\cite{Lentati_2015}, might be better suited to certain types of signals, so we plan to investigate using them in the future.

\subsubsection{Signal model}
\label{ssec:signal_model}

In order to impose the proper coherence between pulsars, in the signal model we model the GW waveform itself as a sum of wavelets, and then project that onto the pulsar lines of sight taking into account the corresponding antenna factors. Since PTAs are sensitive to the time integral of the metric perturbation, we introduce:
\begin{equation}
 H_{+,\times} (t) = \int h_{+,\times} (t) \rm{d} t,
\end{equation}
where $h_{+,\times}$ are the usual components of the metric perturbation corresponding to plus and cross polarized GWs. We model $H_{+,\times} (t)$ as:
\begin{eqnarray} 
H_{+} (t) =  \sum_{j=1}^{N} \Psi(t; t_{0,j}, f_{0,j}, \tau_{j}, A_{+,j}, \phi_{0,+,j}), \\ 
H_{\times} (t) =  \sum_{j=1}^{N} \Psi(t; t_{0,j}, f_{0,j}, \tau_{j}, A_{\times,j}, \phi_{0,\times,j}),
\end{eqnarray}
so that the two polarizations have independent amplitude and initial phase, but $t_0$, $f_0$, and $\tau$ are common parameters. This simplification will not affect the generality of our model, but helps in efficiency by reducing the number of parameters. Then a rotation around the propagation direction is allowed, which introduces the transformation:
\begin{eqnarray} 
\bar{H}_{+} (t) = H_{+} (t) \cos (2 \psi) - H_{\times} (t) \sin (2 \psi) , \\ 
\bar{H}_{\times} (t) = H_{+} (t) \sin (2 \psi) + H_{\times} (t) \cos (2 \psi),
\end{eqnarray}
where $\psi$ is the polarization angle. Then the timing residuals are formed with the use of the antenna pattern functions:
\begin{equation}
 w_{ki}(\bm{\theta}_{\rm s}) = - F_+ (\Omega_k, \Omega_{\rm GW}) \bar{H}_{+} (t_{ki}; \bm{\lambda'}) - F_{\times} (\Omega_k, \Omega_{\rm GW}) \bar{H}_{\times} (t_{ki}; \bm{\lambda'}),
 \label{eq:earth_term_signal}
\end{equation}
where $\Omega_k = (\theta_k, \phi_k)$ and $\Omega_{\rm GW} = (\theta_{\rm GW}, \phi_{\rm GW})$ are the sky location of the $k$th pulsar and the GW source, respectively. $F_{+,\times} (\Omega_k, \Omega_{\rm GW})$ are the antenna pattern functions, and $\bm{\lambda'}$ contains the internal parameters of the GW signal.

Note that eq. (\ref{eq:earth_term_signal}) assumes that only the Earth term (see e.g.~\cite{corbin_cornish2008}) contributes to the signal. This simplification can be justified as follows: The time difference between the Earth and pulsar terms is given by
\begin{equation}
 \Delta t = L (1 - \cos \beta) \approx 50 \ {\rm yr} \left( \frac{ \beta }{10^{\circ}} \right)^2  \left( \frac{L}{1 \ \rm{kpc}} \right),
\end{equation}
where $L$ is the distance to the pulsar, $\beta$ is the angle between the pulsar and the GW source sky location, and for the second equality we made an approximation valid for small $\beta$ (which is where we can get the smallest delays). We can see that for typical millisecond pulsar distances of ~1 kpc and observing campaigns not longer than a few decades, we will not see the pulsar term signals unless the sky location of the GW source is within a few degrees of a pulsar. Even if that happens, it will only appear in that single pulsar, so it can be modeled as a noise transient.

Note that pulsar term signals from GW bursts with no Earth term observed will appear as noise transients in our dataset. For an array with $N_{\rm p}$ pulsars, these correspond to an effective observation time $N_{\rm p}$ times larger than the actual observation time. However, the signal-to-noise ratio (SNR) of a pulsar term signal is a factor of $N_{\rm p}^{1/2}$ lower compared to the Earth term signal, since it appears in just a single pulsar. This smaller SNR corresponds to a decrease in the observable volume by a factor of $N_{\rm p}^{3/2}$. As a result, we expect that the rate of pulsar term bursts will be a factor of $N_{\rm p}^{1/2}$ lower than Earth term bursts, corresponding to a factor of ~7 for the 45 pulsars in the NANOGrav 12.5-year dataset. Pulsar term signals can also be coherent between two (or a few) pulsars, but for an array with a few tens of pulsars, the expected rate of such events are even lower~\cite{pulsar_terms_in_burst_search}.

\subsection{Sampler}
\label{ssec:model}

As described above, the dimensionality of our model depends on whether the GWB is turned on or not, and also on how many wavelets are used in the signal and the transient noise model. A GWB can carry 1 or 2 parameters depending on whether the spectral slope of the power-law model is fixed or varied. Each noise transient wavelet carries 6 parameters including the 5 internal parameters plus an ``external'' parameter indexing which pulsar the noise transient belongs to. The signal model has 3 external parameters common to all wavelets (sky location and polarization angle) plus 7 internal parameters for each wavelet.

To sample this variable-dimension parameter space we use an RJMCMC sampler called \texttt{BayesHopperBurst}, which not only gives posterior distributions on the model parameters, but also on the number of different components included in the model. \texttt{BayesHopperBurst} uses many of the proposals developed for \texttt{BayesHopper} such as Fisher matrix proposals, parallel tempering and trans-dimensional proposals. The description of these can be found in~\cite{BayesHopper}. We also need a global proposal to help convergence, however the one used in \texttt{BayesHopper} is specific to continuous wave sources, so instead we use a global proposal based on the $\tau$-scan, which we describe below.

The $\tau$-scan is defined as:
\begin{equation}
 \mathcal{T} \left( \{ x_i \}; t'_0, f'_0, \tau' \right) = \sum_{i=1}^{N_{\rm p}} |(x_i|\Psi_{c,i})_i|^2 + |(x_i|\Psi_{s,i})_i|^2,
\end{equation}
where $x_i$ is the time series in the $i$th pulsar, $N_{\rm p}$ is the number of pulsars in the array, $(.|.)_i$ is the noise-weighted inner product in the $i$th pulsar, $\Psi_{c,i} = \Psi (A=\bar{A}_i, t_0=t_0', f_0=f_0', \tau=\tau', \phi_0=0)$ is a ``cosine wavelet'' and $\Psi_{s,i} = \Psi (A=\bar{A}'_i, t_0=t_0', f_0=f_0', \tau=\tau', \phi_0=\pi/2)$ is a ``sine wavelet'', where $\bar{A}_i$ and $\bar{A}'_i$ are normalization constants chosen for each pulsar so that $(\Psi_{c,i}|\Psi_{c,i}) = (\Psi_{s,i}|\Psi_{s,i}) = 1$.

By calculating $\mathcal{T}$ for various $t'_0$, $f'_0$, and $\tau'$ values on a grid, we can get a 3D map showing where the data has some excess power. To determine the necessary spacing between grid points so that they properly cover the parameter space, we calculate the match ($M$) between two neighboring wavelets and require that it is above some threshold, e.g.~$M>0.9$. The match between two waveforms ($h$ and $h'$) is defined as:
\begin{equation}
 M = \frac{(h|h')}{\sqrt{ (h|h) (h'|h')}},
\end{equation}
where $(.|.) = \sum_{i=1}^{N_{\rm p}} (.|.)_i$ is the array-wide noise-weighted inner product. $M$ is normalized so that $-1 \leq M \leq 1$, where $M=1$ indicates that $h$ and $h'$ are exactly the same, while $M=-1$ means there is a perfect anticorrelation between them. We can approximate $M$ for two wavelets close to each other as~\cite{BayesWave}:
\begin{equation}
 M = 1 - \frac{1}{4 \bar{\tau}^2} (\Delta \tau)^2 - \frac{1}{2 \bar{\tau}^2} (\Delta t_0)^2 - \frac{\pi^2 \bar{\tau}^2}{2} (\Delta f_0)^2,
\end{equation}
where $\Delta$ indicates the difference in the given parameter between the two wavelets and $\bar{\tau} = (\tau_i + \tau_j)/2$ is the average $\tau$ between the two wavelets.

We can then require that the mismatch ($1-M$) when moving only one of the parameters be less than 0.1. This implies:
\begin{eqnarray}
 \Delta \ln \tau \leq \sqrt{\frac{2}{5}}, \\
 \Delta t_0 \leq \frac{\tau}{\sqrt{5}}, \\
 \Delta f_0 \leq \frac{1}{\sqrt{5} \pi \tau}.
\end{eqnarray}
We see that the $\tau$ layers should be logarithmically spaced with each layer having a $\tau$ value maximum $\exp (\sqrt{2/5}) \approx 1.9$  times larger than the previous one. Time and frequency should be linearly spaced, with the spacing depending on the $\tau$ of the given layer. One can always choose to make the spacing finer, but that results in increasing computational costs.

\begin{figure}[!htb]
 \centering
   \includegraphics[width=0.9\textwidth]{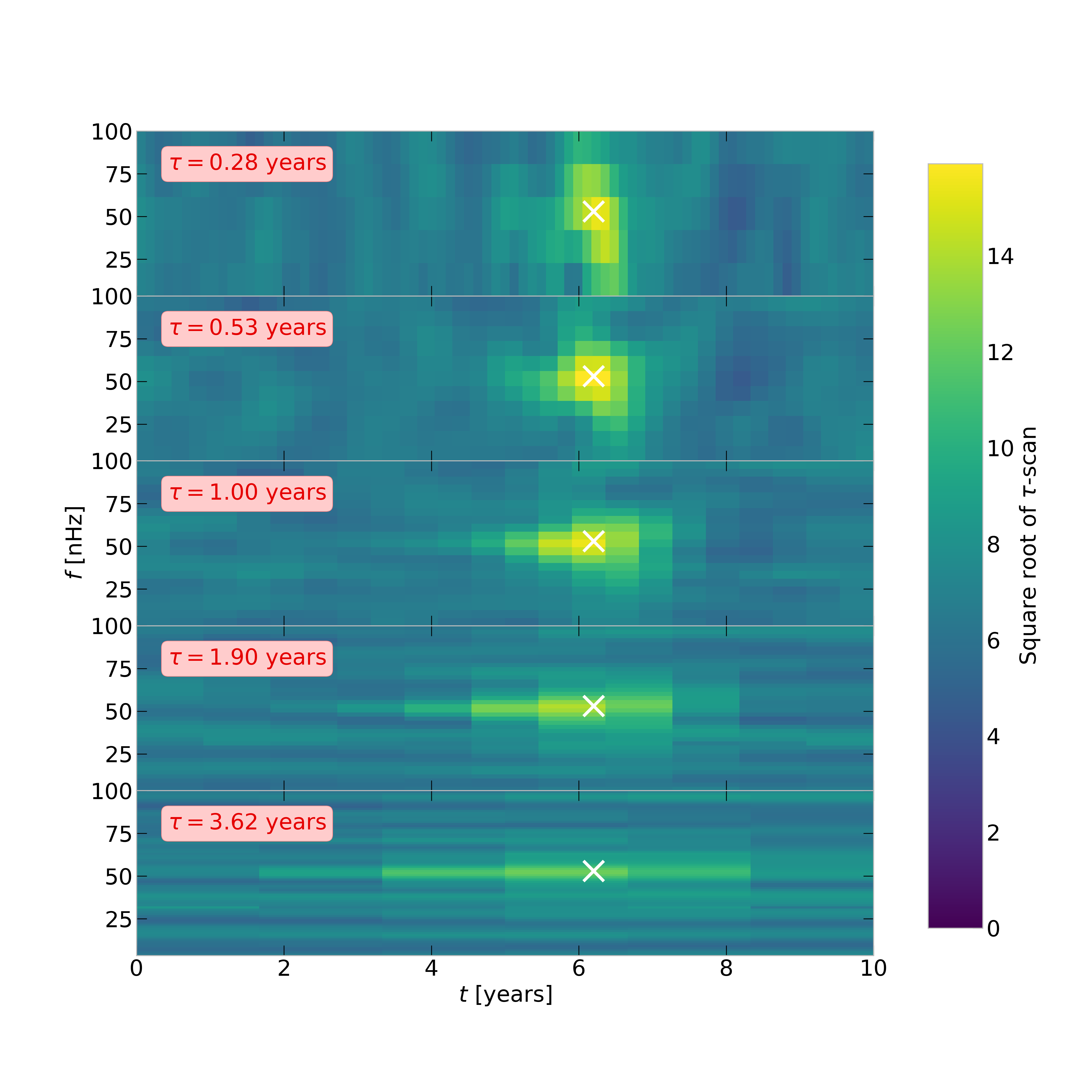}
 \caption{Tau-scan map for a simulated white noise burst signal with central time and frequency marked by the white crosses. Note how the map peaks around the true time-frequency location of the signal, suggesting that such a proposal will help finding the signal quickly.}
 \label{fig:tau_scan_proposal}
\end{figure}

Figure~\ref{fig:tau_scan_proposal} shows the $\tau$-scan calculated for a white noise burst signal (see Section~\ref{ssec:signal_only}) with central time and frequency marked with the white crosses. Here we use 5 layers with different $\tau$ values. We can see that as a result of the $\tau$-dependent time and frequency spacing, pixels at higher $\tau$ values are more elongated in time but narrower in frequency (resulting in the same time-frequency area). We can see that the $\tau$-scan clearly lights up around the injected signal indicating that, after proper normalization, this provides an excellent proposal. Indeed, we find that including this proposal speeds up convergence by multiple orders of magnitude, even though the remaining parameters of the wavelet are drawn from their priors. This proposal is also used when proposing to add a new signal wavelet, which makes it possible to have a reasonable acceptance rate for these complicated jumps. 

A variant of the $\tau$-scan proposal is used for noise transient wavelets. For these we create a $\tau$-scan for each pulsar only using the data for the pulsar in question. These are not only used to draw $f_0$, $t_0$ and $\tau$ values for the wavelet, but also to draw which pulsar to put in the wavelet based on how much excess power is in each pulsar.

\section{Tests on simulated data}
\label{sec:injection_tests}

To test \texttt{BayesHopperBurst} we analyze a few different simulated datasets. All the datasets in this section correspond to a simulated PTA with 20 pulsars at random sky locations. Each has been observed for 10 years every 30 days, has the same constant white noise level of 0.5 $\mu$s and no red noise. We add different kinds of GW signals and noise transients to this base dataset. In Section~\ref{ssec:signal_only} we show results for datasets only containing GW signals and not noise transients; in Section~\ref{ssec:glitch_only} we look at a scenario with no GW signals but noise transient in many pulsars of the array; while in Section~\ref{ssec:signal_and_noise} we look at a few examples of how \texttt{BayesHopperBurst} can deal with datasets containing both GW signals and noise transients.

\subsection{Signal without noise transients}
\label{ssec:signal_only}

\begin{figure}[!htb]
 \centering
   \includegraphics[width=0.85\textwidth]{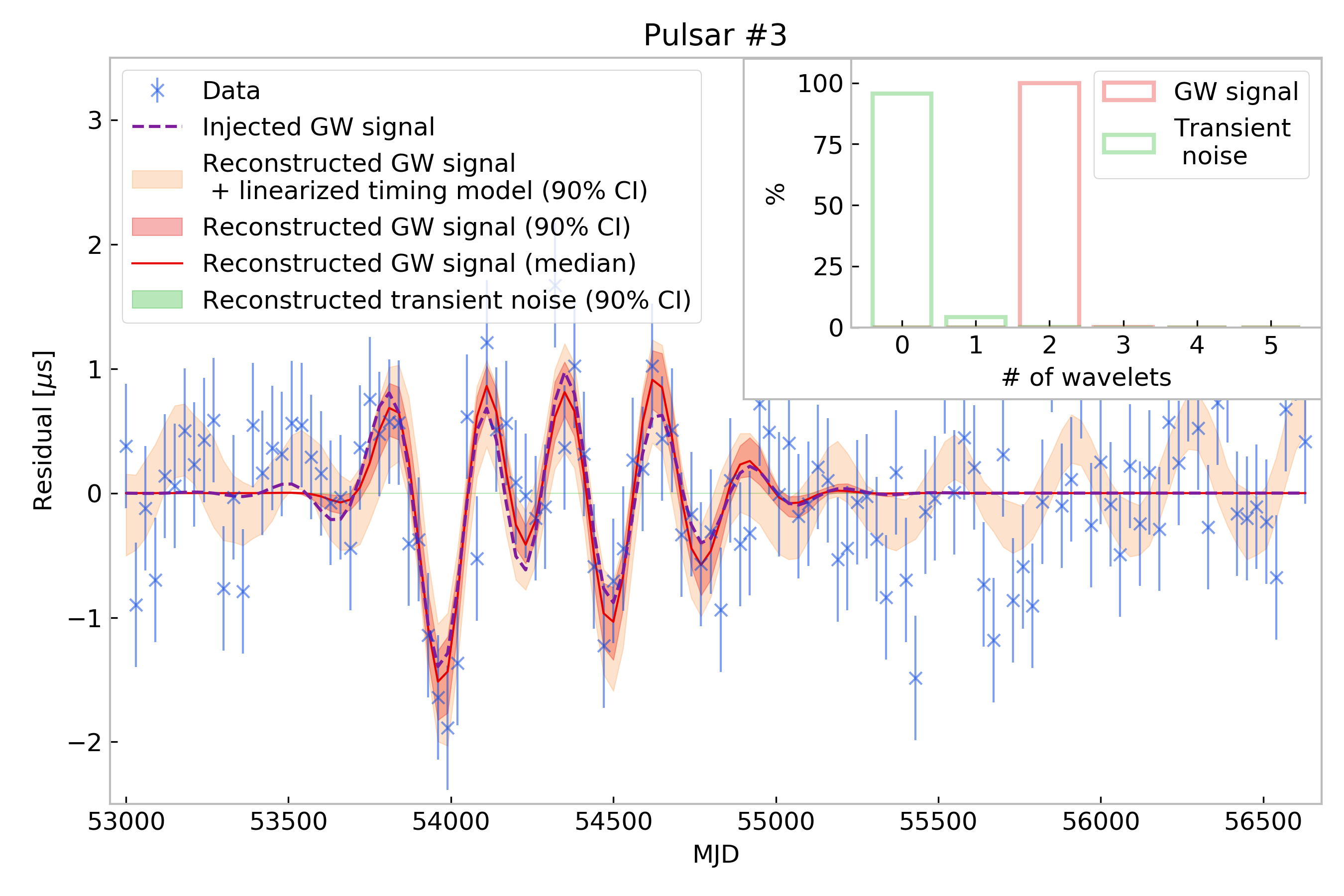}
 \caption{Reconstruction of a signal consisting of two sine-Gaussian wavelets (${\rm SNR}=17.9$) in one of the pulsars in the array. The reconstruction is more precise than one would expect by looking at the spread of data points because the reconstruction is based on data from all 20 pulsars in the array. The match between the median reconstructed waveform and the injected waveform is 0.96. The inlet plot shows the histogram of the number of wavelets used both in the signal and the noise transient model. We can see that \texttt{BayesHopperBurst} uses two signal wavelets and no noise transients as expected.}
 \label{fig:multiple_sg_reconstruction}
\end{figure}

First we look at a dataset with a GW signal that consists of two sine-Gaussian wavelets that are close to each other in the time-frequency plane. This can be considered as one of the easiest signals to reconstruct, since it matches our model, i.e., we can perfectly reconstruct such a signal with two signal wavelets. Figure~\ref{fig:multiple_sg_reconstruction} shows the reconstruction of this signal in one of the pulsars in the array. We can see that \texttt{BayesHopperBurst} uses no noise transient wavelets and two signal wavelets most of the time, as expected. We can also see that the reconstructed signal matches well the injected GW signal. Note that while Figure~\ref{fig:multiple_sg_reconstruction} only shows the reconstruction in a particular pulsar, the projection of the signal in the datastream of other pulsars can be significantly different. The reconstructions for all 20 pulsars in the array are shown in Figure~\ref{fig:multiple_sg_wallpaper}, indicating similarly accurate waveforms in all pulsars. To quantitatively characterize the goodness of fit throughout the whole array, we calculate the match between the median reconstructed signal and the injected signal, which we find to be 0.96. The SNR of the simulated signal is 17.9, which we define as ${\rm SNR}=\sqrt{(h_{\rm inj}|h_{\rm inj})}$, where $h_{\rm inj}$ is the waveform of the simulated (or injected) signal.

The next dataset contains a white noise burst (WNB), which is produced from a white noise time series by applying a Gaussian window both in time and frequency. The variance of the white noise sets the amplitude of our WNB, while the windowing determines its time-frequency location and spread. While a WNB is similar to a sine-Gaussian wavelet in that it is also well localized on the time-frequency plane, it has a more complicated time dependence. This also means that a WNB cannot be perfectly reproduced by a finite number of sine-Gaussian wavelets, so the number of wavelets used will be determined by the trade-off between getting an increasingly better fit and paying an Occam penalty for each additional wavelet included.

\begin{figure}[!htb]
 \centering
   \includegraphics[width=0.85\textwidth]{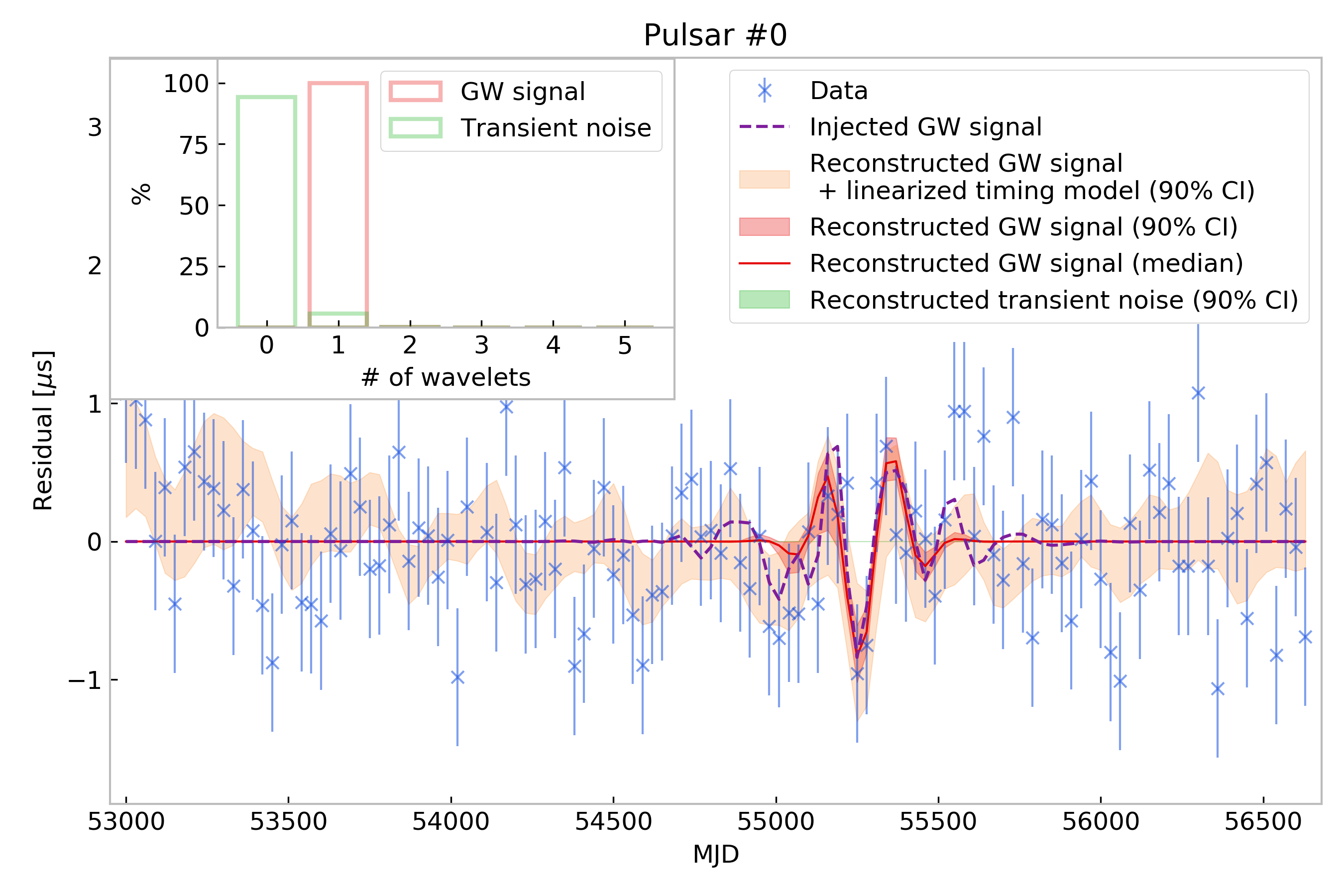}
 \caption{Reconstruction of a white noise burst signal (${\rm SNR}=18.3$) in one of the pulsars of the array. The match between the median reconstructed waveform and the injected waveform is 0.812. We can see that parts of the signal are modeled by the linearized timing model. The inlet plot shows the histogram of the number of wavelets used both in the signal and the noise transient model. We can see that \texttt{BayesHopperBurst} uses a single signal wavelet to model the WNB.}
 \label{fig:wnb_reconstruction}
\end{figure}

Figure~\ref{fig:wnb_reconstruction} shows the reconstruction of a WNB signal. We can see that \texttt{BayesHopperBurst} uses a single wavelet to fit the majority of the signal, while low-amplitude parts of the signal are missed. This also leads to a lower match between the median reconstructed signal and the injected signal (0.812) compared to the previous example, even though it has a slightly higher SNR of 18.3. This suggests that \texttt{BayesHopperBurst} performs slightly worse for a WNB signal compared to sine-Gaussians. Note however that if we take into account the uncertainty in the linearized timing model (orange band on Figure~\ref{fig:wnb_reconstruction}), the reconstruction is consistent with the injected GW signal at the 90\% confidence level.

One of the potential sources of GW bursts in the PTA band are SMBHBs on parabolic or highly eccentric orbits. Here we analyze the simulated signal from a parabolic encounter of two SMBHs. We calculate the waveform from eq.~(4-3) of~\cite{Finn_Lommen2010}. Beyond being a potential astrophysical source, parabolic SMBHBs are also a good test of the algorithm, because they are not elliptically polarized, so \texttt{BayesHopperBurst} must use its flexibility to fit two very different waveforms for the plus and cross polarizations.

\begin{figure}[!htb]
 \centering
   \includegraphics[width=0.85\textwidth]{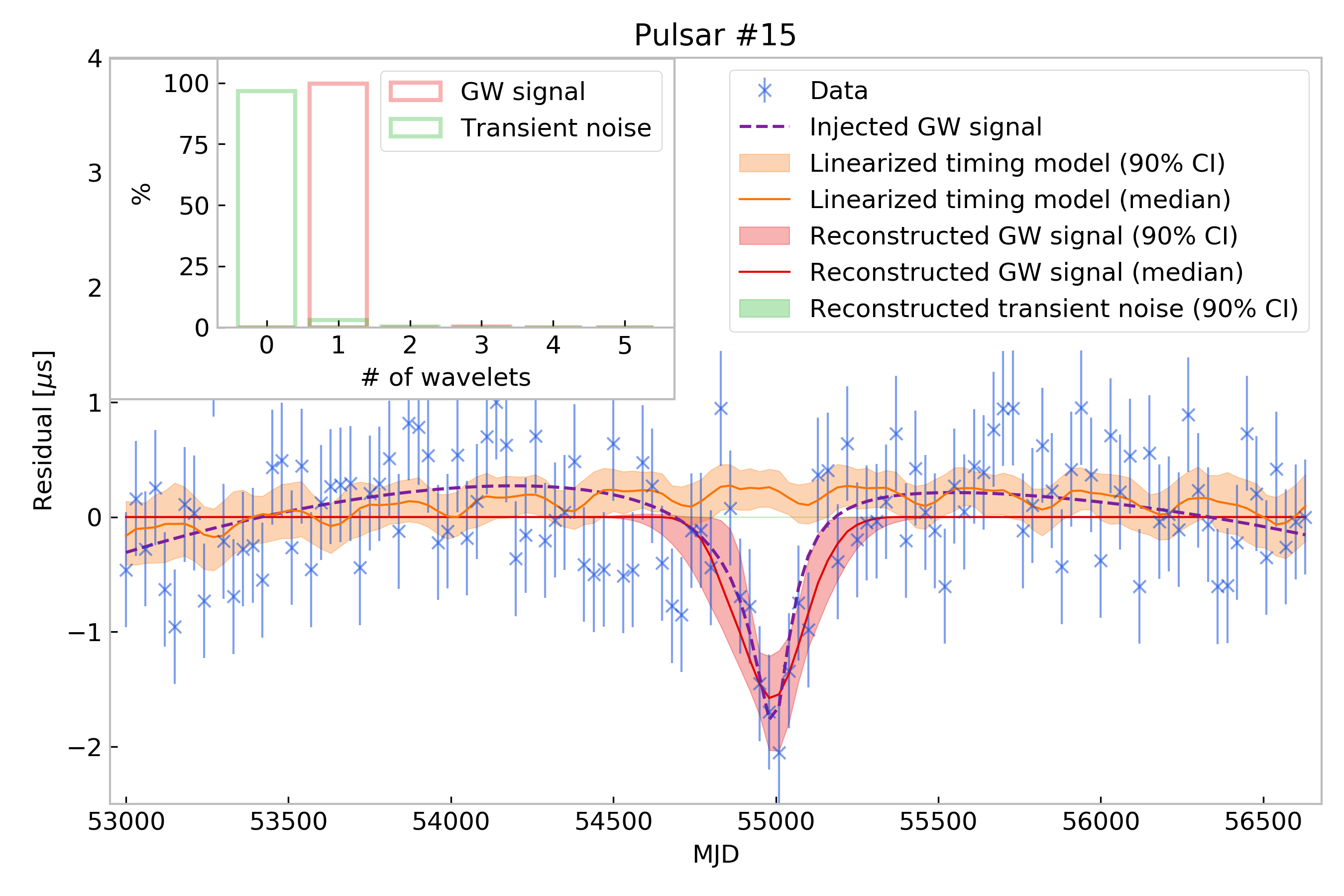}
 \caption{Reconstruction of a parabolic BBH signal (${\rm SNR}=16.8$) in one of the pulsars of the array. The BHs have equal masses of $10^9 M_{\odot}$, and are on an orbit with an impact parameter of 180 $M_{\odot}$ at a distance of 15 Mpc. The match between the median reconstructed waveform and the injected waveform is 0.956. We can see that the low-frequency part of the signal is modeled by the linearized timing model, while the high-frequency part is reconstructed using a single signal model wavelet, as can be seen on the inlet plot showing the histogram of the number of wavelets used.}
 \label{fig:parabolic_bbh_reconstruction}
\end{figure}

Figure~\ref{fig:parabolic_bbh_reconstruction} shows the waveform reconstructed by \texttt{BayesHopperBurst} for an equal mass SMBHB on a parabolic orbit with a total mass of $M = 2 \times 10^9 M_{\odot}$, an impact parameter of $b = 60 M$, at a distance of 15 Mpc, with an SNR of 16.8. Unlike wavelets or WNBs, these parabolic encounter waveforms have a significant low-frequency component. We can see that the linearized timing model can fit this component with a quadratic function by changing the period and period derivative parameters of the timing model. The high-frequency (and high-amplitude) part of the signal is reconstructed by a single wavelet. The match between the median reconstructed signal and the injected signal is 0.956.

\subsection{Multiple noise transients}
\label{ssec:glitch_only}


We have seen that \texttt{BayesHopperBurst} is able to reconstruct GW bursts in the presence of Gaussian noise. In this section, we test if it can correctly identify incoherent noise transients. We consider four different datasets: i) with a single pulsar containing a sine-Gaussian noise transient; ii) with 3 pulsars containing the same noise transient; iii) with 10 pulsars (half the array) containing the same noise transient; and iv) all 20 pulsars containing the same noise transient. Having the same noise transient in many pulsars can be considered a worst case scenario for discriminating signals and noise transients. It is virtually impossible to have such a situation by chance alignment of incoherent noise transients, but it provides a good test case. Note that even if all the pulsars have the exact same noise transient, in principle, that can be distinguished from a coherent GW burst, because there is no sky location and polarization that would result in the exact same waveform (with the same amplitude) in each pulsar.

\begin{figure}[!htb]
 \centering
   \includegraphics[width=0.7\textwidth]{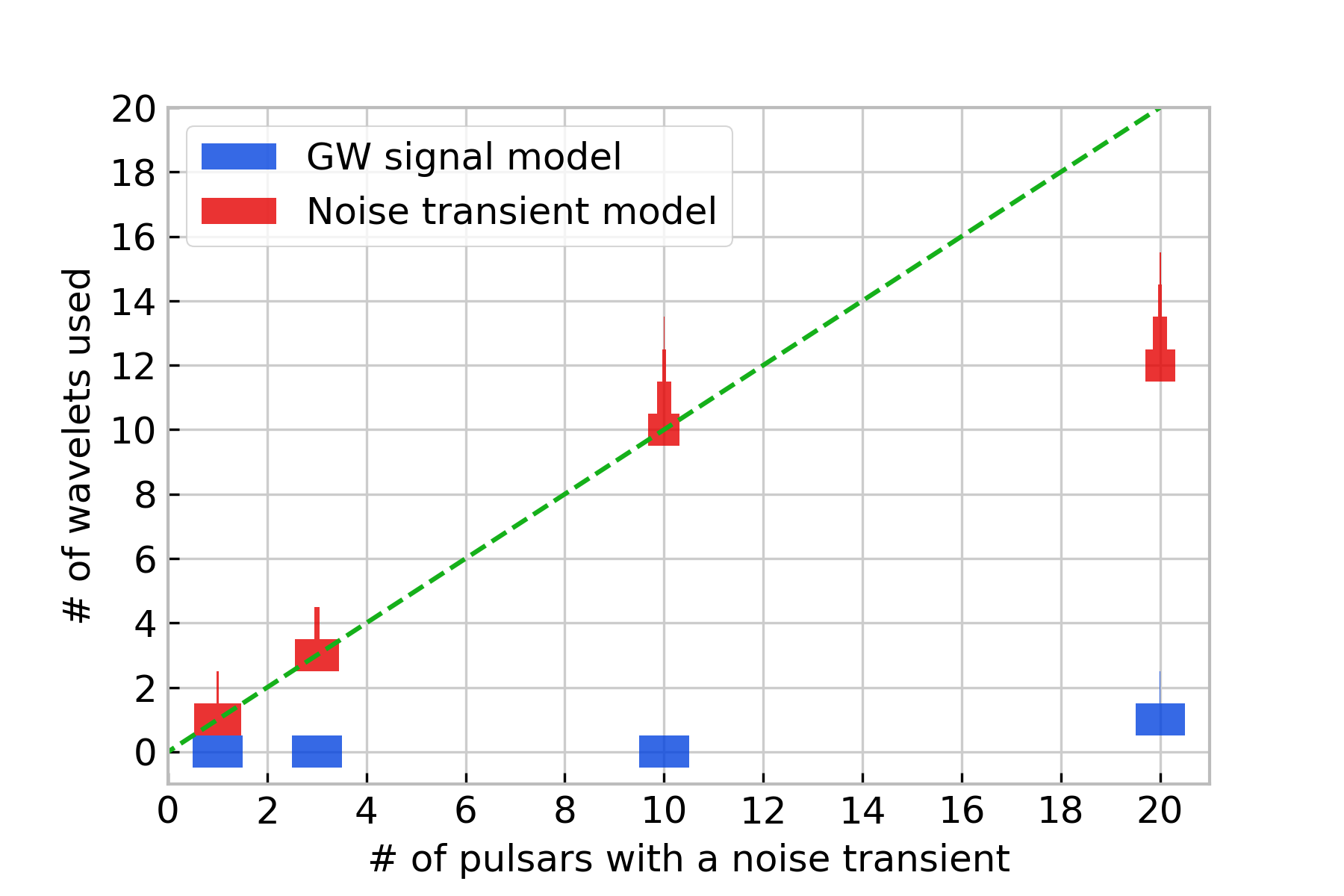}
 \caption{``Discrete violin plot'' of number of wavelets used in signal and noise transient model for datasets with the exact same noise transient present in a different number of pulsars. We can see that when the noise transient is present in not more than half of the pulsars, the model is correctly identified. However, when the noise transient is present in all the pulsars, the sampler instead turns on a signal model and finds a sky location which takes care of most of the signal and fits the rest with noise transient wavelets.}
 \label{fig:glitch_only_model_dimension}
\end{figure}

Figure~\ref{fig:glitch_only_model_dimension} shows the histogram of the number of wavelets used in the signal model and the noise transient model for each of these datasets. We can see that for the first three datasets we are recovering the correct model of having 1/3/10 noise transient wavelets and no signal wavelets. For the dataset with all the pulsars containing the same noise transient we instead see that \texttt{BayesHopperBurst} is using a single signal wavelet and 12 to 15 noise transient wavelets. This can be understood as a result of the inbuilt parsimony of Bayesian statistical analysis. As noted above, there are no such external parameters that could result in the exact same amplitude in the datastream of all the pulsars. However, one can use the signal model to account for noise transients in a few pulsars, and use noise transient wavelets to model the rest. A single-wavelet signal model has 10 parameters, while each noise transient wavelet has 5 parameters. Thus such a solution could be statistically favored if the signal model can account for noise transients in more than two pulsars. Looking at the number of noise transient wavelets used, we can see that in this example the signal model is able to mimic the residuals from up to 8 noise transients.

\subsection{Signal and noise transients}
\label{ssec:signal_and_noise}

\begin{figure}[!htbp]
 \centering
    \begin{subfigure}[b]{0.8\textwidth}
            \centering
            \includegraphics[width=\linewidth]{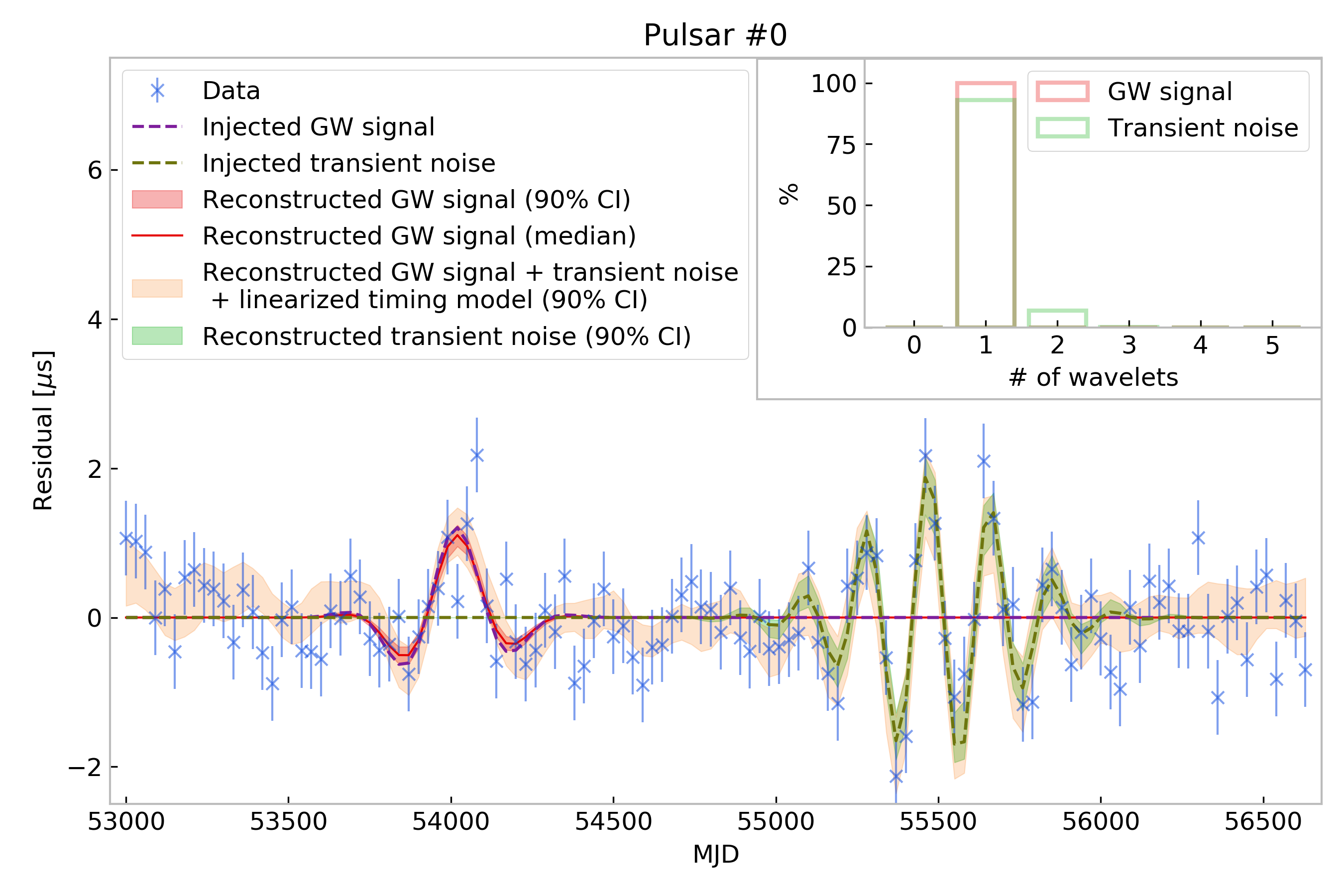}
            \caption{}
    \end{subfigure}%
    \hfill
    \begin{subfigure}[b]{0.8\textwidth}
            \centering
            \includegraphics[width=\linewidth]{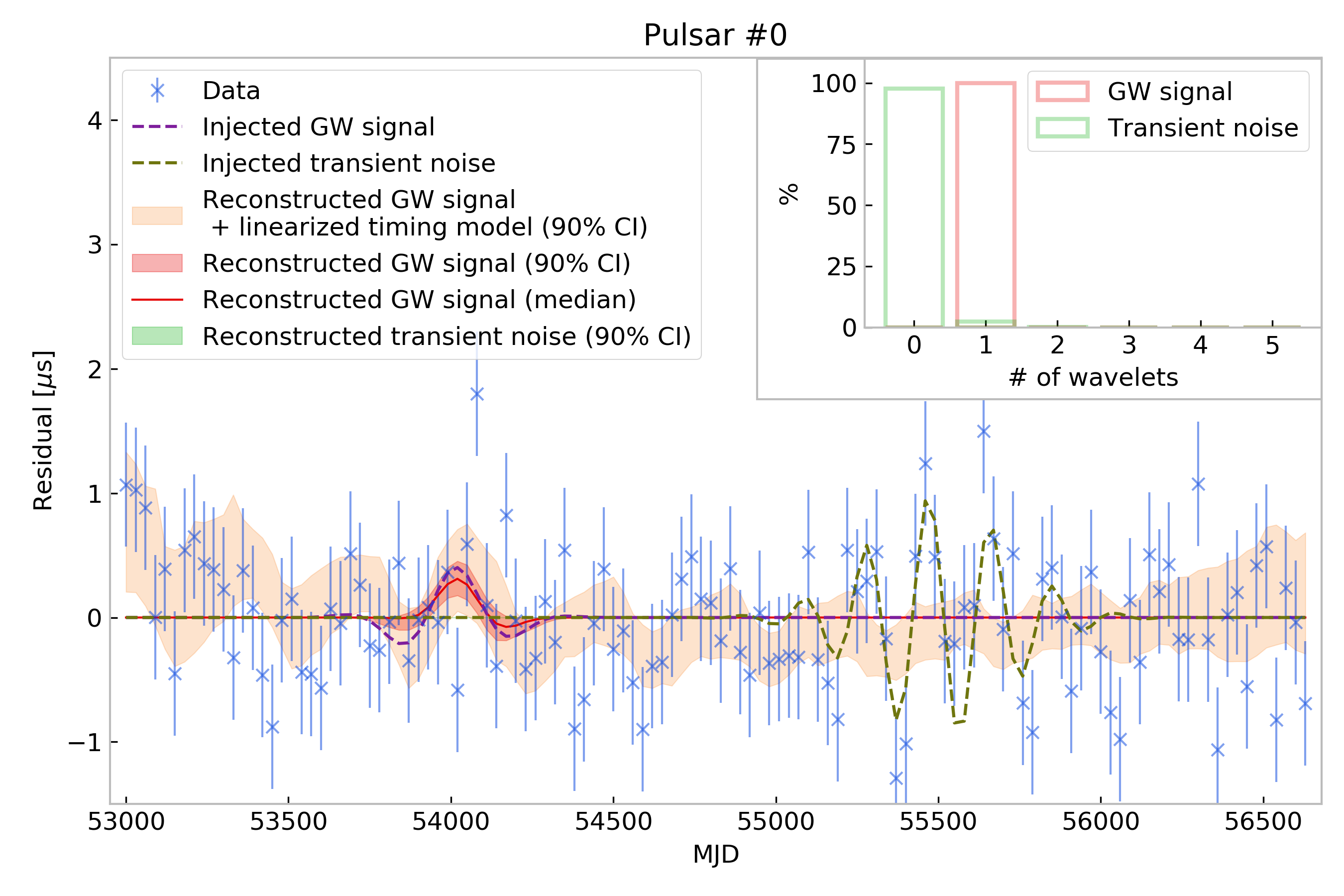}
            \caption{}
    \end{subfigure}%
    \hfill
 \caption{Reconstruction of signal and noise transient well separated in time. In the case of high amplitudes (a), both signal (${\rm SNR}=26.8$) and noise transient (${\rm SNR}=10$) are accurately reconstructed resulting in a match of 0.994 between the median reconstructed and injected waveforms. In case of low amplitudes (b), only the signal (${\rm SNR}=8.9$) is reconstructed by \texttt{BayesHopperBurst} with $M=0.87$, while the noise transient (${\rm SNR}=5$) is found to be insignificant. This is not surprising given the low SNR of the noise transient.}
 \label{fig:signal_plus_glitch_well_separated}
\end{figure}

We have seen how \texttt{BayesHopperBurst} can reconstruct signals or noise transients. In this section we test it on datasets containing both signals and noise transients to see how it is able to distinguish between the two. First we analyze a dataset containing a signal and a noise transient in one of the pulsars, both of which have sine-Gaussian waveforms but with significantly different $t_0$, $f_0$ and $\tau$ parameters. We produce two versions of this dataset with different amplitudes: i) with signal SNR of 26.8, and noise transient SNR of 10; ii) with signal SNR of 8.9, and noise transient SNR of 5.  The reconstructions for these datasets are shown in Figure~\ref{fig:signal_plus_glitch_well_separated}. We can see that both the signal and the noise transients are correctly identified and reconstructed for the higher amplitude case, and we find $M=0.994$ between the injected and recovered waveforms. For the lower amplitude case only the signal has been found with a moderate match of 0.87. Note however that both the relatively lower match value and the fact that no noise transient has been found are consistent with the significantly lower SNRs of both signal and noise transient in this dataset.



\begin{figure}[!htbp]
 \centering
    \begin{subfigure}[b]{0.6\textwidth}
            \centering
            \includegraphics[width=\linewidth]{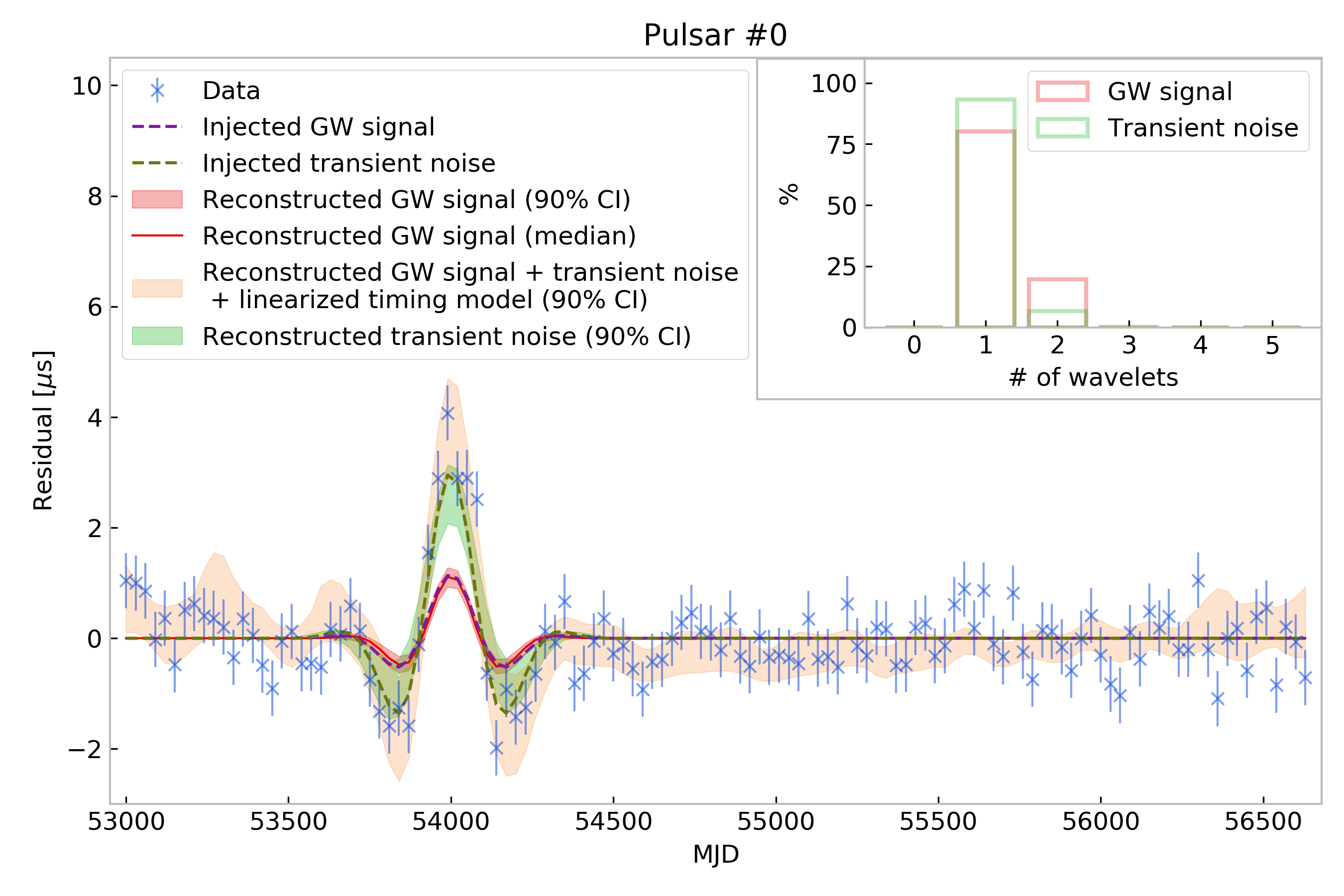}
            \caption{}
            \label{fig:worst_case_scenario1_reconstruction_1}
    \end{subfigure}%
    \hfill
            \begin{subfigure}[b]{0.4\textwidth}
            \centering
            \includegraphics[width=\linewidth]{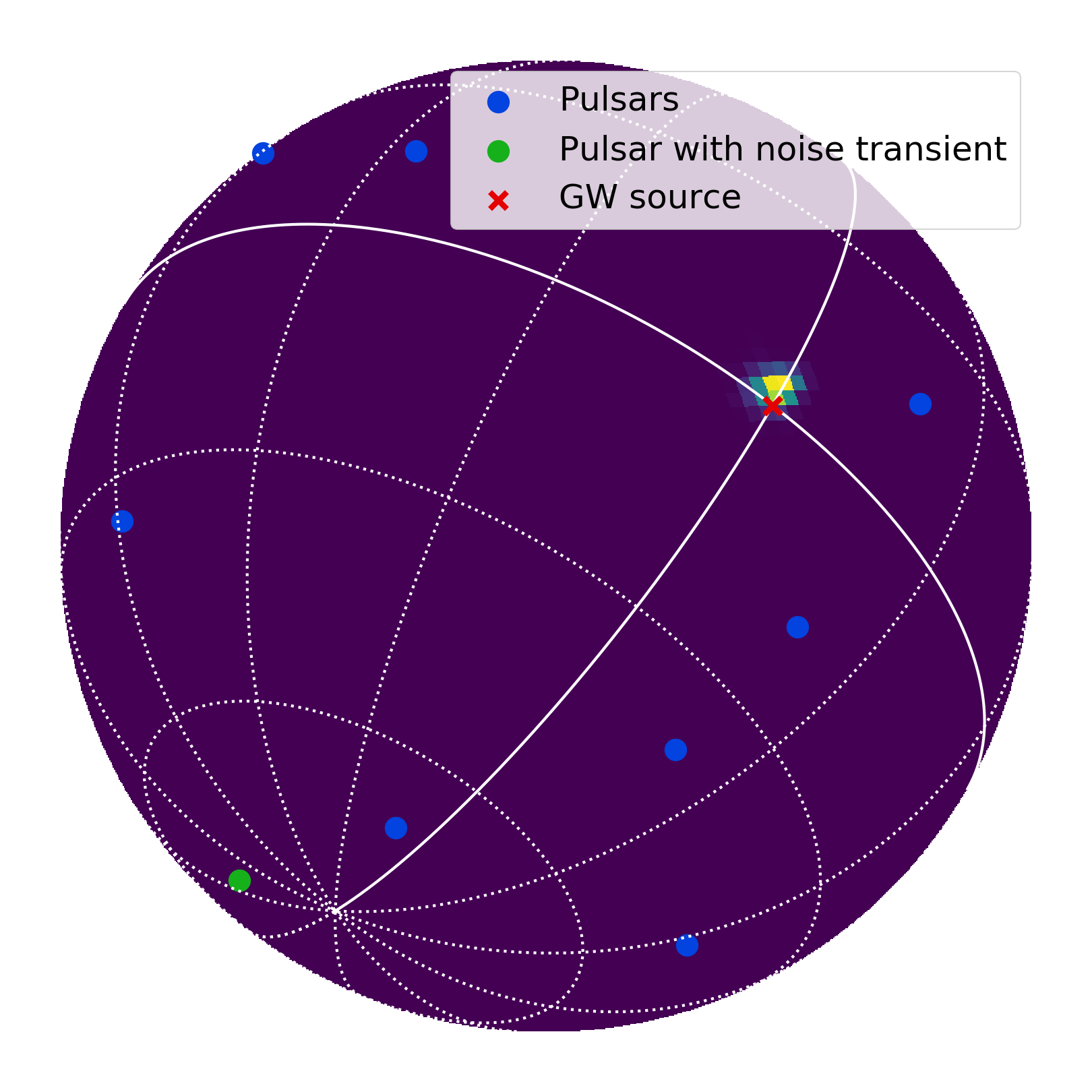}
            \caption{}
            \label{fig:worst_case_scenario1_skymap_1}
    \end{subfigure}%
    \hfill
    \begin{subfigure}[b]{0.6\textwidth}
            \centering
            \includegraphics[width=\linewidth]{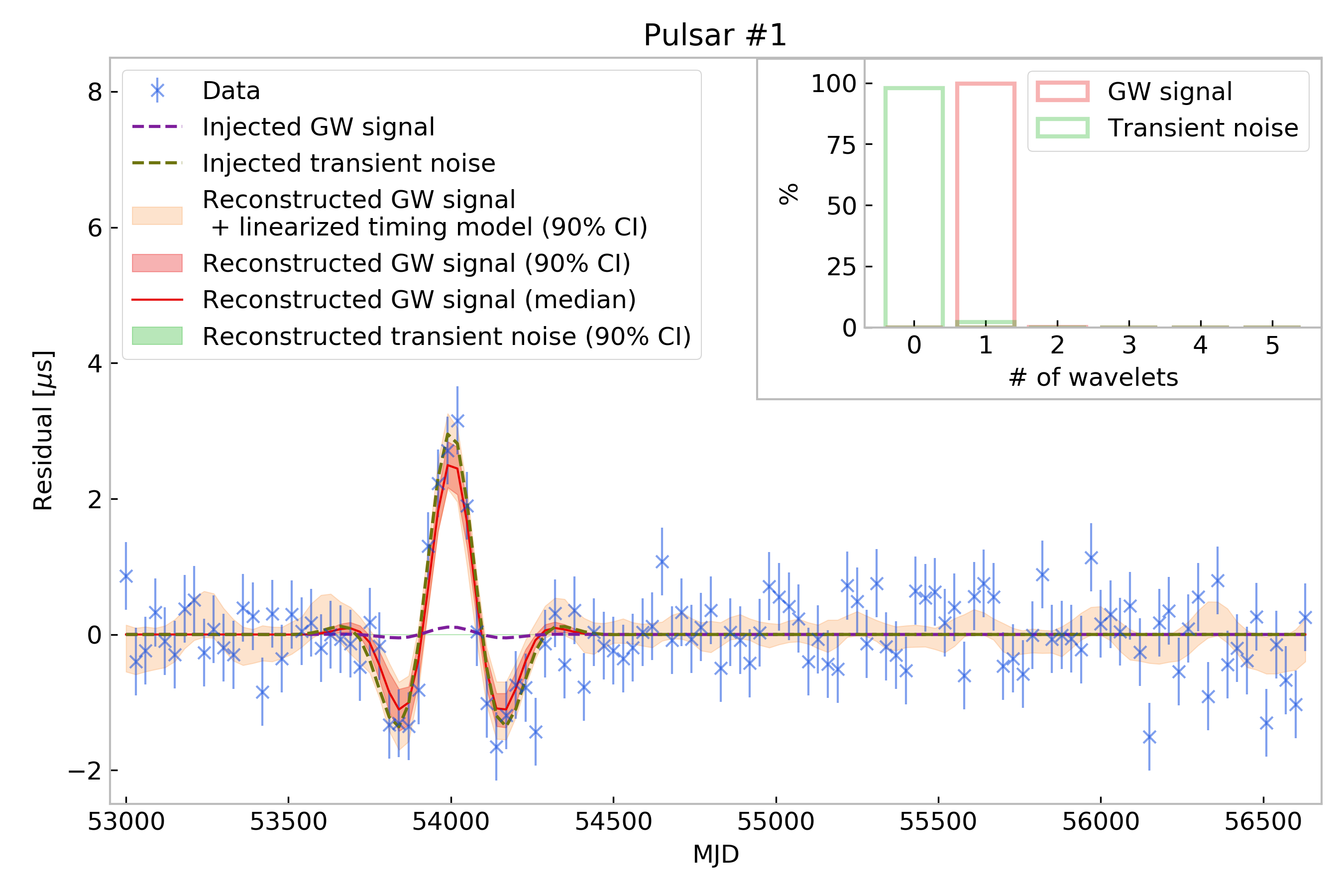}
            \caption{}
            \label{fig:worst_case_scenario1_reconstruction_2}
    \end{subfigure}%
    \hfill
        \begin{subfigure}[b]{0.4\textwidth}
            \centering
            \includegraphics[width=\linewidth]{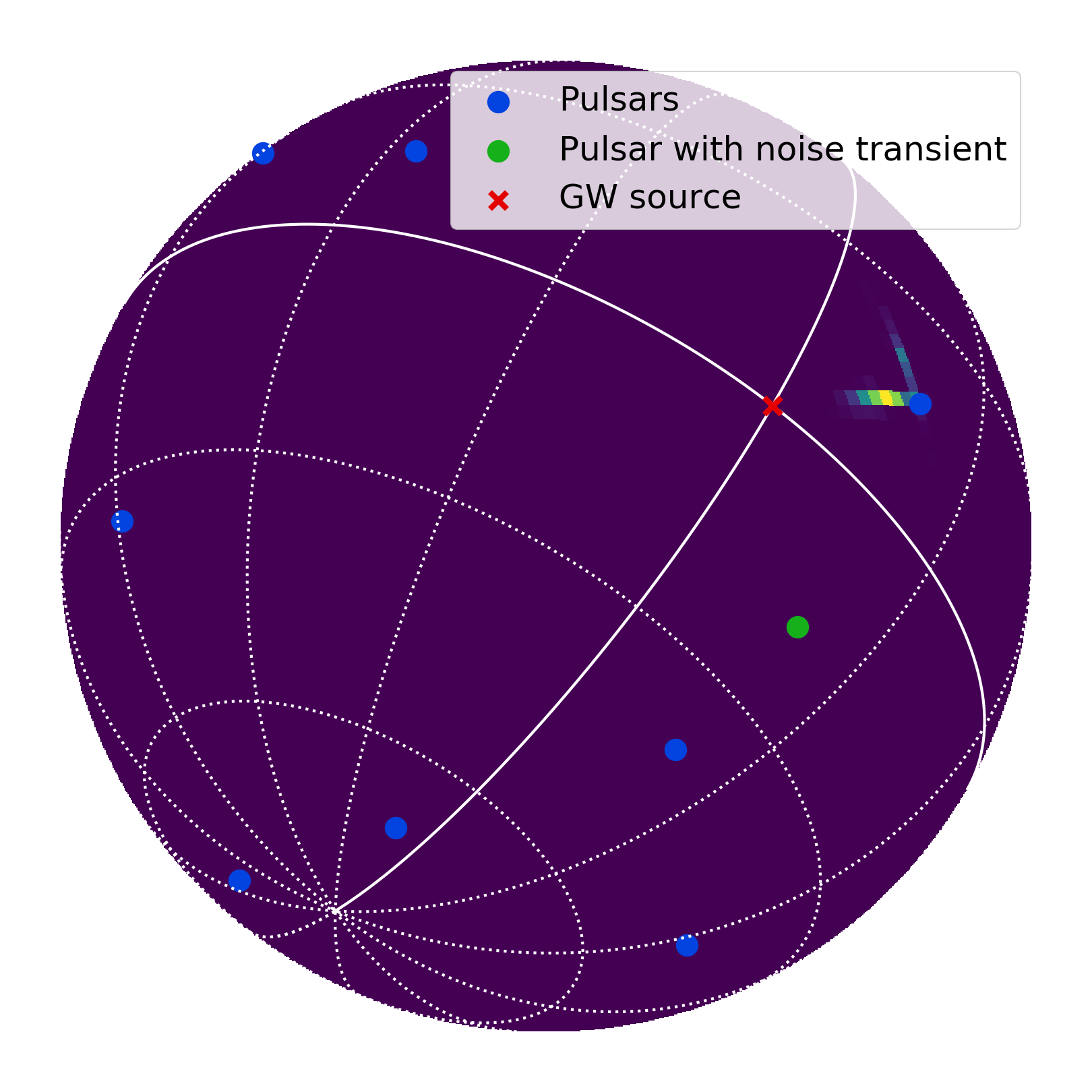}
            \caption{}
            \label{fig:worst_case_scenario1_skymap_2}
    \end{subfigure}%
    \hfill
 \caption{Reconstruction of a dataset with a sine-Gaussian GW signal (${\rm SNR}=16.1$) and a noise transient (${\rm SNR}=10.8$) with the exact same waveform. Both noise transient and signal are correctly reconstructed when the noise transient is in a pulsar far away from the GW source on the sky (a-b). However, when the noise transient is in a pulsar close to the GW signal location, the effect of the noise transient can be accounted for by slightly changing the sky location of the GW without affecting the signal in other pulsars (c-d).}
 \label{fig:worst_case_scenario1_reconstruction}
\end{figure}

In the next example, we create a dataset with a sine-Gaussian signal (${\rm SNR}=16.1$) and a noise transient (${\rm SNR}=10.8$) in one of the pulsars with the exact same waveform as the signal. This can be considered a worst case scenario, since \texttt{BayesHopperBurst} can only rely on the coherence of the signal when trying to distinguish GW signal and noise transient. We make two versions of this dataset with the noise transient being in two different pulsars. Figure~\ref{fig:worst_case_scenario1_reconstruction_1} shows the reconstruction for the case where the noise transient is in a pulsar far away from the GW source on the sky as shown by the sky location posterior map (see Figure~\ref{fig:worst_case_scenario1_skymap_1}). We can see that both the waveforms and the sky location is accurately reconstructed in this case ($M=0.98$). However, having a noise transient in a pulsar that lies close to the true GW source sky location results in no significant noise transient reconstructed (see Figure~\ref{fig:worst_case_scenario1_reconstruction_2}). Instead, \texttt{BayesHopperBurst} fits the data by changing the sky location slightly to account for the effect of the noise transient (see corresponding skymap in Figure~\ref{fig:worst_case_scenario1_skymap_2}). Since the noise transient is in one of the pulsars closest to the GW source on the sky, the location of the GW source can be changed slightly in a way that it can account for the noise transient in that pulsar but not change significantly the projection of the signal in other pulsars. This reconstruction results in a match of 0.96, which is only slightly lower than the previous example, even though it is not using the true model.



\section{Upper limit predictions for the NANOGrav 12.5-year dataset}
\label{sec:upper_limit}
In a follow-up study, we plan to use \texttt{BayesHopperBurst} to search for nHz GW bursts in the NANOGrav 12.5-year dataset~\cite{nanograv_12p5yr_data}. To assess the sensitivity we might expect from such an analysis we run \texttt{BayesHopperBurst} on a simulated dataset made to resemble the NANOGrav 12.5-year dataset and used in~\cite{Astro4Cast}. The dataset has been made using the properties (timing parameters, white and red noise parameters) of pulsars and observational time stamps from the NANOGrav 12.5-year dataset. Note that the red noise parameters are derived by filtering out a common red noise process, so that they truly characterize pulsar intrinsic red noise. To reduce computational costs the residuals are epoch averaged. More details on this simulated dataset can be found in~\cite{Astro4Cast}.

\begin{figure}[htb]
 \centering
   \includegraphics[width=0.9\textwidth]{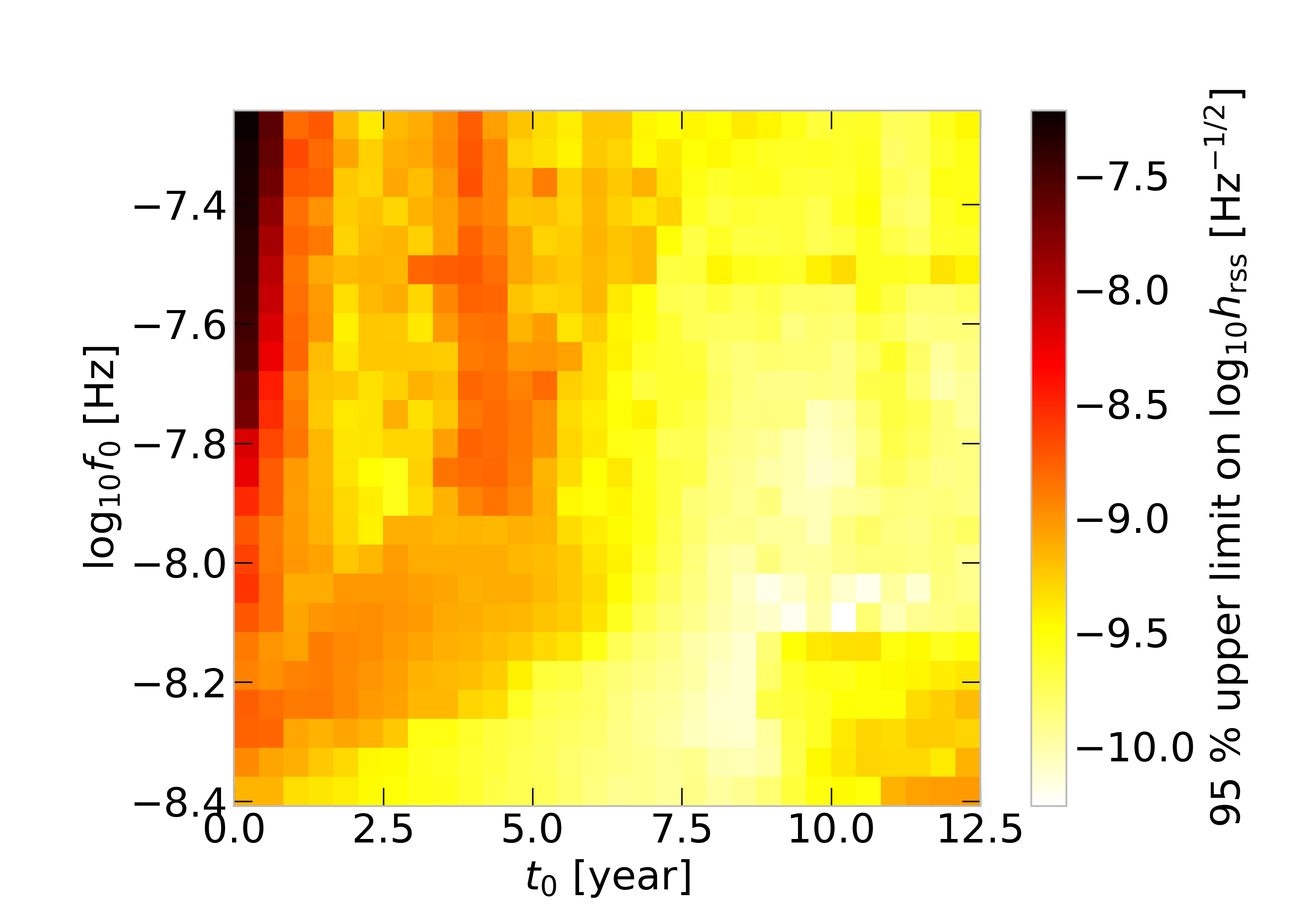}
 \caption{Upper limit on the root-sum-squared strain amplitude ($h_{rss}$) as a function of central time ($t_0$) and central frequency ($f_0$) based on a simulated dataset made to resemble the NANOGrav 12.5-year dataset~\cite{Astro4Cast}. The most sensitive time-frequency location is at around 10 years -- 8 nHz where we get an upper limit of $h_{rss} \simeq 5 \times 10^{-11}$ Hz$^{-1/2}$, which corresponds to $\sim 40 M_{\odot} c^2$ emitted in GWs at a fiducial distance of 100 Mpc.}
 \label{fig:upper_limit_test}
\end{figure}

This dataset does not contain any GW bursts, so by running \texttt{BayesHopperBurst} on it we can set upper limits on the amplitude of generic GW transients. We consider these upper limits to be indicative of the upper limits we might expect to get from running on the real NANOGrav 12.5-year dataset assuming we do not find any significant GW candidates. Figure~\ref{fig:upper_limit_test} shows the uppper limits we get from this simulated dataset as a function of central time ($t_0$) and central frequency ($f_0$) defined e.g.~by equations (8a) and (8b) in~\cite{BayesWavePE}, which coincide with the parameters used to describe the wavelets if only 1 wavelet is used. We quote these upper limits in terms of the root-sum-squared strain amplitude defined as:
\begin{equation}
 h_{rss} = \sqrt{  \int_{-\infty}^{\infty} \left[ h_+^2 + h_{\times}^2 \right] \rm{d}t } = \sqrt{  \int_{-\infty}^{\infty} \left[ \left( \frac{\rm{d} H_+}{\rm{d} t} \right)^2 + \left( \frac{\rm{d} H_{\times}}{\rm{d} t} \right)^2 \right] \rm{d}t }.
\end{equation}
To produce this plot we calculate $t_0$, $f_0$ and $h_{rss}$ for each sample from our MCMC. Then we bin the samples based on a grid on the $t_0$--$f_0$ plane, and in each bin we calculate the 95th percentile of the $h_{rss}$ samples from that bin, which we quote as the 95\% upper limit.

A clear trend from Figure~\ref{fig:upper_limit_test} is that the upper limit decreases as we move to later times. This is because the array contains more and more pulsars as time goes on, which greatly improves the array's sensitivity. In addition, we see drastic improvements in the second half of the dataset, which can be explained by the improved backends used after $\sim$7.5 year and a high-cadence observing campaign, which started at $\sim$8.5 year (see e.g.~\cite{nanograv_12p5yr_data}). We can also see that the sensitivity decreases at the temporal edges of the dataset, which is due to the fact that waveforms with a central time near the edges will have a significant portion that lies outside of the observational time span of the dataset. The worst sensitivity in the analyzed time-frequency region is achieved at very high frequencies and very early times. This is due to the fact that there are less than 10 pulsars observed at those times. There is also a slight increase in the upper limit around the frequency corresponding to a period of 1 year ($\log_{10} f_0\simeq-7.5$), where PTAs are known to be less sensitive due to the sky location fitting in the timing model.

The best sensitivity is achieved around $t_0 \simeq 10$ years and $f_0 \simeq 8$ nHz at an amplitude of $h_{rss} \simeq 5 \times 10^{-11}$ Hz$^{-1/2}$. This can be converted into the minimum GW energy that can be detected at a fiducial distance using e.g.~equation (4.2) of~\cite{all_sky_initial_ligo}. We get that at a fiducial distance of 100 Mpc we are sensitive to sources emitting an energy of more than $\sim 40 M_{\odot} c^2$ in GWs.

Note that we use a uniform prior on the wavelet amplitude which pushes up the amplitude values resulting in a conservative upper limit. However, as we also allow the sky location to change, and the distribution of the pulsars is highly anisotropic, this also pushes the sampler to the least sensitive sky locations where the highest amplitude signals can remain hidden. This means that the sensitivity at the best sky location could be significantly better than the value quoted above. On the other hand, in this run we fix the red noise parameters in each pulsar to the true values due to computational constraints. This in principle could make our upper limit estimates overly optimistic as it does not take into account any covariances between the red noise and GW burst signal models. However, since the red noise only dominates at the lowest frequencies, we do not expect this to have a drastic effect at the most sensitive time-frequency location reported above. In addition, even while fixing the red noise parameters to their true values, we still allow the realization of the red noise to vary, which can accommodate some covariance between the signal and red noise models.

\section{Single-pulsar test on real data}
\label{sec:real_data_test}
\begin{figure}[!htbp]
 \centering
    \begin{subfigure}[b]{0.9\textwidth}
            \centering
            \includegraphics[width=\linewidth]{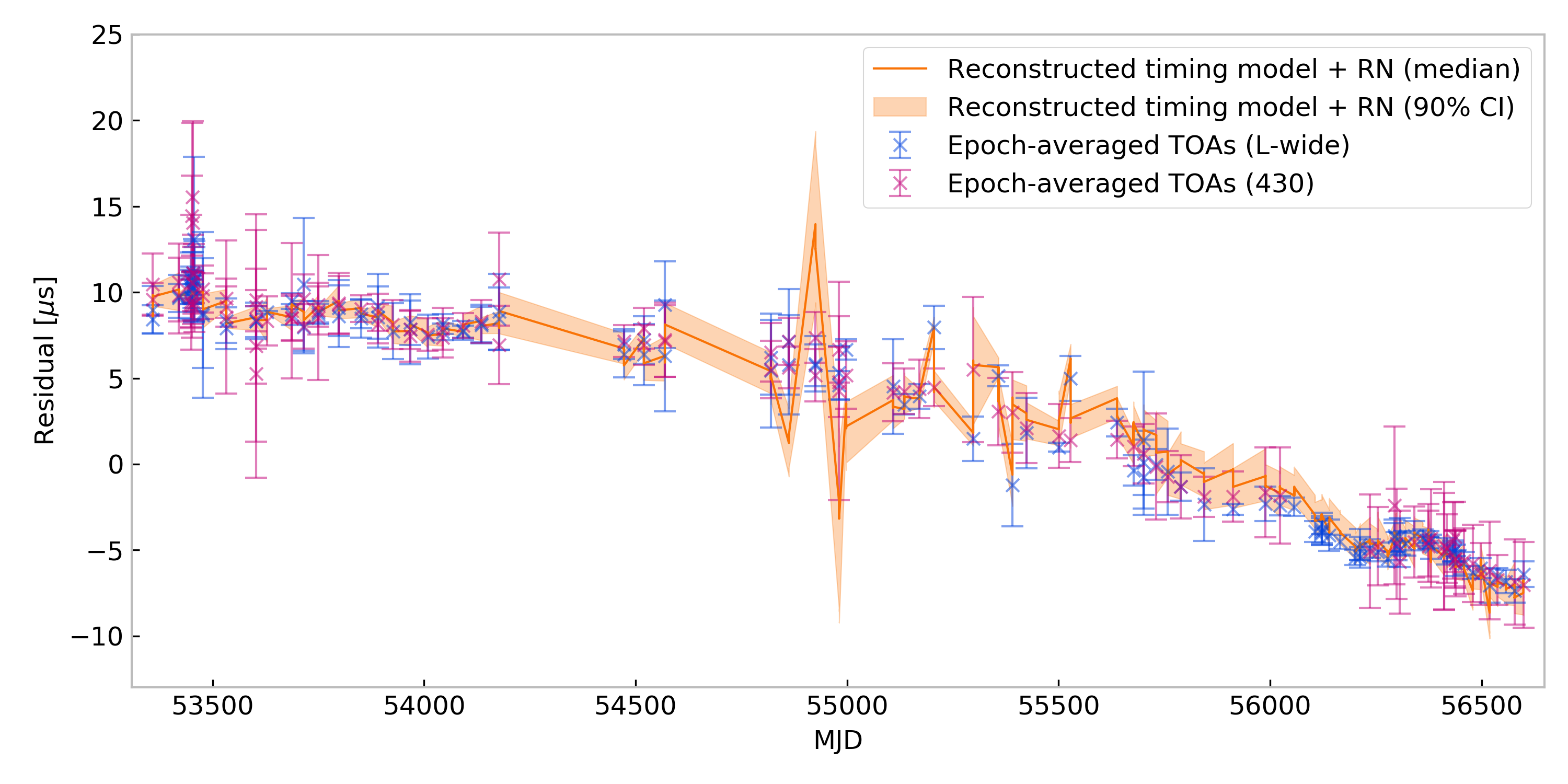}
            \caption{}
            \label{fig:9yr_test_a}
    \end{subfigure}%
    \hfill
    \begin{subfigure}[b]{0.9\textwidth}
            \centering
            \includegraphics[width=\linewidth]{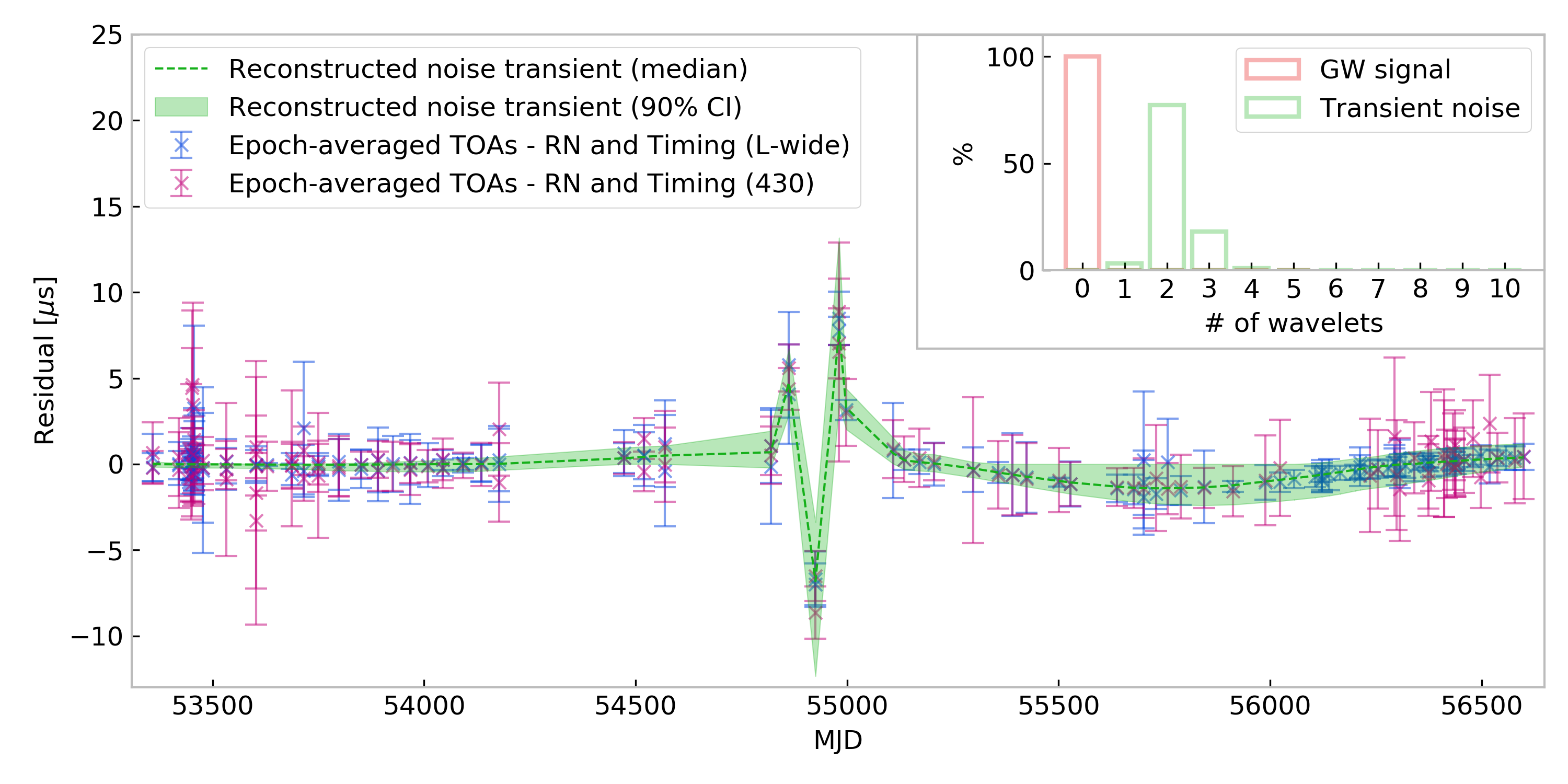}
            \caption{}
            \label{fig:9yr_test_b}
    \end{subfigure}%
    \hfill
 \caption{Waveform reconstruction for B1855+09 from the NANOGrav 9-year dataset. (a) shows the original epoch-averaged residuals along with the reconstructed timing model and red noise model realizations. (b) shows the epoch-averaged residuals corrected by the median red noise and timing model realizations, along with the noise transient reconstruction from \texttt{BayesHopperBurst}. The latter suggests that there is a noise transient in this dataset around MJD 55000. Note that there is a strong covariance between the presence of the noise transients and DMX parameters.}
 \label{fig:9yr_test}
\end{figure}

Real data can bring complications compared to simulated data like uneven sampling, temporal gaps in the datastream, chromatic effects, etc. To test how \texttt{BayesHopperBurst} can handle these complications we run it on a single pulsar (B1855+09) from the NANOGrav 9-year dataset~\cite{nanograv_9yr_data}. We show the epoch-averaged residuals from B1855+09 in Figure~\ref{fig:9yr_test_a} along with the reconstruction of the linearized timing model and the red noise model. Figure~\ref{fig:9yr_test_b} shows the epoch averaged residuals corrected with the median timing and red noise model. These indicate a transient feature around 55000 Modified Julian Date (MJD), which is reconstructed by the noise transient model also shown on this figure. Note that although we show the epoch-averaged residuals for plot clarity, our analysis used all individual residuals.

 We can see in Figure~\ref{fig:9yr_test} that \texttt{BayesHopperBurst} adjusts the timing parameters such that some features of the data were smoothed out while others were amplified. Then it used a few wavelets to correct those amplified features. In particular, we can see that the observations made at different frequencies at the same epoch are sometimes inconsistent in the original dataset. One can correct that by changing the value of the dispersion measure parameter (DMX, see e.g.~\cite{nanograv_9yr_data}) for that epoch, but that will result in a net residual in both observations. The legacy analysis thus cannot do that, since it cannot account for transient noise features. On the other hand, \texttt{BayesHopperBurst} can change the dispersion measure values and model the resulting noise transient with a collection of wavelets. Thus, our analysis suggests that there is a noise transient in this dataset. Such a noise transient was not found before, since the canonical analysis does not model noise transients and their presence can be masked by inflating the white noise parameters. In fact, \texttt{BayesHopperBurst} only finds the noise transient when we allow the white noise parameters to vary, so that it can lower their values compared to the canonical analysis.
 
 Based on Figure~\ref{fig:9yr_test}, it might look like that the feature around 55000 MJD fitted by the noise transient model was introduced by the similar feature with opposite sign in the timing and red noise model. Note however that the linearized timing model introduces perturbations around timing model parameters previously determined by an analysis which does not allow for unmodeled noise transients. Such an analysis will try to adjust its free parameters to model out any noise transients present in the dataset. Our results indicate that if we introduce a model for noise transients, the Bayesian analysis prefers to undo that timing model adjustment, and instead model the feature with the dedicated noise transient model.

\section{Conclusion}
\label{sec:conclusion}
We have presented \texttt{BayesHopperBurst}, a Bayesian algorithm to search for nHz GW burst using PTA data, by modeling them as a collection of sine-Gaussian wavelets, where the number of wavelets is found by a trans-dimensional RJMCMC sampler. We demonstrated how it can reconstruct signals with a wide range of morphologies, and also how it can distinguish between coherent GW burst and incoherent noise transients that may occur in the datastream of some pulsars. 

In the near future, we plan to use \texttt{BayesHoperBurst} to carry out the first ever search for generic GW bursts in the nHz frequency regime using the NANOGrav 12.5-year dataset. In preparation for that we analyzed a simulated dataset similar to the NANOGrav 12.5-year dataset and made predictions on the sensitivity we might expect from such a search. To demonstrate that \texttt{BayesHopperBurst} is capable of dealing with the additional complications arising when analyzing real data, we analyzed a single pulsar (B1855+09) from the NANOGrav 9-year dataset.

In addition to searching for generic GW bursts, an algorithm like \texttt{BayesHopperBurst} could also be a useful tool for analyzing transient noise features in the datastream of single pulsars as illustrated in~\cite{Justin_Neil_transdim_noise}. We are planning to further explore the potential of such an approach in a follow-up study. Systematically analyzing a large collection of pulsars will also help us better understand the covariances we have seen in Section \ref{sec:real_data_test} between transient noise, DMX, and red noise models.

Important areas to investigate in future studies are further development and optimization of \texttt{BayesHopperBurst}. One particularly interesting direction would be experimenting with different types of functions in our signal and noise transient models (e.g.~shapelets~\cite{Lentati_2015}), which might be more suited for certain signals or noise transients. 

\ack
The authors thank David Nice and Justin Ellis for informative discussions about B1855+09, and Sarah Vigeland for feedback on the manuscript.
We appreciate the support of the NSF Physics Frontiers Center Award PFC-1430284.
Some computations were performed on the Hyalite High Performance Computing System, operated and supported by University Information Technology Research Cyberinfrastructure at Montana State University.

\appendix

\section{Waveform reconstruction in all pulsars}
\label{sec:appendix}

\begin{figure}[htb]
 \centering
   \includegraphics[width=1.0\textwidth]{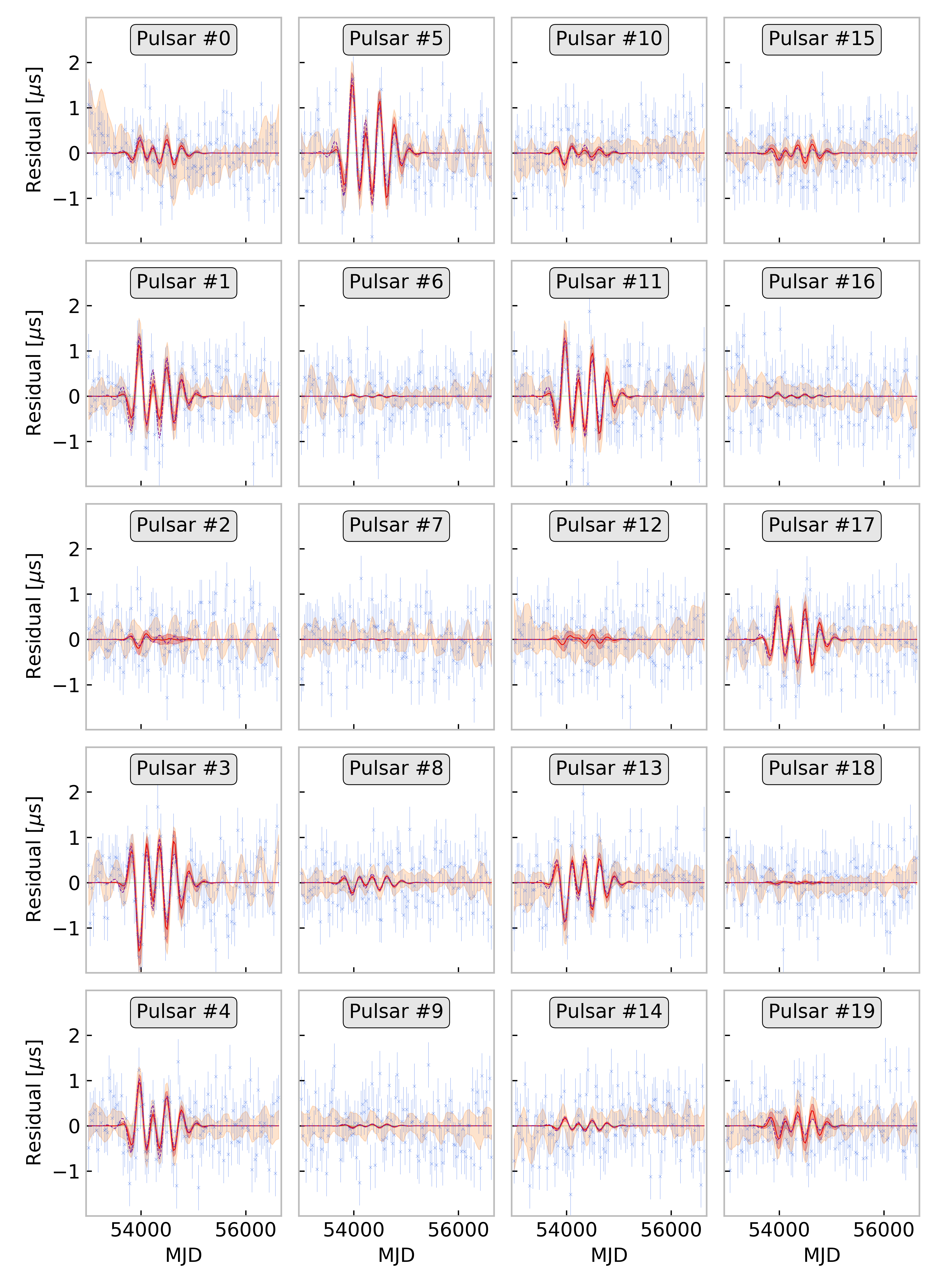}
 \caption{Reconstruction of a signal consisting of two sine-Gaussian wavelets in all 20 pulsars in the array. This result is from the same dataset for which the reconstruction in pulsar \#3 is shown in Figure~\ref{fig:multiple_sg_reconstruction}. For more details see the caption of Figure~\ref{fig:multiple_sg_reconstruction}.}
 \label{fig:multiple_sg_wallpaper}
\end{figure}

In Section~\ref{sec:injection_tests}, we only show the waveforms reconstructed in a single pulsar (see Figures~\ref{fig:multiple_sg_reconstruction},~\ref{fig:wnb_reconstruction},~\ref{fig:parabolic_bbh_reconstruction},~\ref{fig:signal_plus_glitch_well_separated}, and~\ref{fig:worst_case_scenario1_reconstruction}), but it is also interesting to look at how the signal appears in different pulsars. Figure~\ref{fig:multiple_sg_wallpaper} shows the simulated data and the reconstructed signal in all 20 pulsars of the array for the injection with two sine-Gaussian wavelets (cf.~Figure~\ref{fig:multiple_sg_reconstruction}). While the amplitude and waveform of the signal appearing in datastreams of different pulsars are significantly different, \texttt{BayesHopperBurst} consistently reconstructs the signal in each pulsar. This is not surprising given that we employ a coherent signal model which uses information from all the pulsars to constrain the waveform of the signal.

\section*{References}
\bibliography{bayeshopperburst_bib}{}
\bibliographystyle{unsrt_et_al}

\end{document}